\documentclass[conference]{IEEETran}

\usepackage{xspace}
\usepackage{xcolor}

\usepackage{setspace}
\usepackage{fancyhdr,graphicx}
\usepackage[ruled,vlined]{algorithm2e}
\SetAlgoCaptionLayout{raggedright}
\usepackage{url}

\begin{document}

\title{Let's shake on it: Extracting secure shared keys from Wi-Fi CSI}

\author{\IEEEauthorblockN{Tomer Avrahami} \IEEEauthorblockA{School of Electrical Engineering,\\ Tel Aviv University, ISRAEL\\ Email: tomer.avrm@gmail.com} \and 
\IEEEauthorblockN{Ofer Amrani} \IEEEauthorblockA{School of Electrical Engineering,\\ Tel Aviv University, ISRAEL\\ Email: ofera@tauex.tau.ac.il} 
\and 
\IEEEauthorblockN{Avishai Wool} \IEEEauthorblockA{School of Electrical Engineering,\\ Tel Aviv University, ISRAEL\\ EMail: yash@eng.tau.ac.il}
}



\maketitle

\begin{abstract}
A shared secret key is necessary for encrypted communications. Since Wi-Fi relies on OFDM, we suggest a method to generate such a key by utilizing Wi-Fi's channel state information (CSI). CSI is typically reciprocal but very sensitive to location: While the legitimate Alice and Bob observe the same CSI, an eavesdropper Eve observes an uncorrelated CSI when
positioned over 0.5 wavelength away.

We show that 
if endpoint Bob is shaken, sufficient diversity is induced in the CSI so that it can serve as a source of true randomness. Then we show that the CSI among neighboring sub-carriers is correlated, so we select a small set of judiciously-spaced sub-carriers, and use a majority rule around each. We demonstrate that Alice and Bob observe a 5-15\% bit mismatch rate (BMR) in the extracted bitstream while Eve observes a BMR of around 50\%  even when placed within 10cm of Alice.

We employ the cryptography-oriented definition of min-entropy to estimate the number of secure bits within the bitstream, and use the Cascade algorithm of quantum-key-distribution to reconcile Alice and Bob's bitstreams, while quantifying the number of bits leaked by the algorithm. Accounting for both the min-entropy and the cascade leakage we quantify the Secured Bit Generation Rate of our method.

We conducted extensive tests 

in an indoor environment.
Our system exhibits a secure bit generation rate of 1.2--1.6 
bits per packet, at distances ranging from 0.5m--9m, and can generate a secure shared 128-bit key with 20sec of device shaking.

\end{abstract}

\section{Introduction}
\subsection{Motivation}
Secured Wi-Fi networks, whose access-points employ security measures such as WPA2 or WPA3 \cite{alliance2019wpa3} require a Preshared Secret encryption key (PSK) that is shared among the access point and all the mobile devices on the network. Typically the PSK is manually configured into the access point, and shared with all the mobile devices either out-of-band or via WPS~\cite{alliance2020wps}. Thus the compromise of the key on a single mobile device could lead to a compromise of the whole Wi-Fi network. 

In this paper we demonstrate the feasibility of a more secure architecture, in which every mobile device has a separate, randomly-generated key, shared only with the access point---without the  logistical challenges of manually sharing multiple separate keys. 

Since Wi-Fi relies on Orthogonal Frequency Division Multiplexing (OFDM), we suggest to generate the PSK based on Wi-Fi's channel state information (CSI). The CSI is reciprocal but very sensitive to location, so while the access point Alice and the mobile device Bob observe the same CSI, an eavesdropper Eve will observe an uncorrelated CSI as long as she is outside the spatial correlation distance---approximately 7cm for Wi-Fi at 2.4GHz band and 3.5cm at 5GHz band.

\subsection{Method Overview}

The goals of our system design are as follows:
\begin{itemize}
    \item The CSI source should be treated as a True Random Bit Generator (TRBG) and measured as such, using cryptographically-accepted metrics such as the min-entropy to estimate the resulting key's cryptographic strength.
    \item The extracted key needs to be agreed-upon by Alice and Bob. This precludes using common TRBG practices like using the least-significant measurement bits to enhance randomness: doing so will create an excessive bit-mismatch-rate (BMR)~due~to~channel~and measurement~noise.
    \item The extracted key must be private from~the~eavesdropping~adversary~Eve.
\end{itemize}

To achieve these goals we use the following procedure:

\begin{enumerate}
    \item Devices Alice and Bob exchange $N$ packets while the user shakes device Bob, and both extract the CSI from the packets they receive. Shaking the device changes the channel between Alice and Bob and introduces randomness into the CSI, making it a suitable source for a TRBG. 
    \item We observed that adjacent sub-carriers have correlated CSI. Therefore Alice and Bob use a carefully-selected set of $k$ well-separated sub-carriers, and quantize their CSI into $q$ bits per sub-carrier, to obtain a raw bitstream (Alice and Bob perform this extraction separately). 
    \item Alice and Bob run the Cascade algorithm \cite{brassard1994secret} to reconcile the raw bitstream: technically Bob reconciles his bitstream to match Alice's. 
    \item Alice and Bob compute a cryptographic hash, e.g., SHA-256 \cite{nist-sha}, on the reconcilled bitstream to compress all the secure bits into a key that can serve as the PSK between Alice and Bob. 
\end{enumerate}

When analyzing the cryptographic strength of the resulting key we make the following observations. First, the Cascade algorithm in step (3) leaks information, but it precisely quantifies the number of bits it leaks. Therefore the number of Cascade-leaked bits is subtracted from the total number of bits collected in step (2) as a basic bound $B_0$ on the number of secure information bits. Moreover, as the $k\cdot q$-wide bit patterns extracted from packets in step (2) are temporally correlated, we calculate their min-entropy and use it to conservatively reduce $B_0$, to get a bound $B$ on the number of remaining secure information bits embedded in the bit stream. As long as $B$ is larger than the hash length then the PSK has the cryptographic strength that its length implies. The value $B$, normalized by the number of packets $N$, is the Secure Bit Generation Rate (SBGR), measured in bits per packet.

As a concrete example, in many experiments we used $N=300$ packets sent in each direction. Via extensive testing we found that using $k=4$ sub-carriers and $q=2$ bits per sub-carrier (i.e., extracting 8 bits per packet) yields a good working point, so the size of the raw bitstream is 2400 bits. 
In this setting, Cascade leaked $\approx 1000$ bits, and we measured a min-entropy of $\approx 2.7$ out of 8, leaving $B\approx 472$ secure bits in the reconciled bitstream, well above the PSK length of 256, and providing SBGR=1.6 secure bits per packet.

Note that it is not informative to run empirical tests of randomness such as the NIST test suite~\cite{nist800-22-1a} on the resulting PSK: such tests would only measure the effectiveness of the cryptographic hash function---the output of a good hash function like SHA-256 will always pass all such tests, regardless of its input.

\subsection{Adversary Model}
We assume that the adversary Eve is passive, and does not attempt to disrupt the key agreement protocol. Eve is also assumed to be equipped with radio receivers that are as sensitive as, or better than, those of Alice and Bob. Thus Eve can eavesdrop on the CSI-bearing packets and try to estimate the CSI characterizing the channel between Alice and Bob. However we assume that Eve is outside the spatial correlation distance $D_c$ from both Alice and Bob: $D_c = 0.5\lambda$ where $\lambda$ is the wavelength \cite{tse2005fundamentals}. For Wi-Fi at 2.4GHz ($\lambda=12.5$cm) we have $D_c=6.25$cm and for Wi-Fi at 5GHz ($\lambda=6$cm) $D_c=3$cm. Consequently, as long as Eve is at least 7cm away from both Alice and Bob, the corresponding channels she experiences, in typical settings, are decorrelated from the channel between Alice and Bob.

As Bob is continuously moving, we assume it would be impossible for Eve to imitate this movement. So, the best location Eve can be in is as close to Alice as possible. In our experiments, we placed Eve~10cm from Alice, which is the Raspberry Pi case's width.

It is assumed that Eve is physically positioned within the Wi-Fi network coverage of both Alice and Bob, i.e., she is in the same room as Alice and Bob. Eve can therefore eavesdrop on all the key reconciliation messages exchanged by the Cascade algorithm~\cite{brassard1994secret} without any errors. Since these messages are sent \emph{before} a cryptographic key is agreed-upon, they are sent in cleartext, so any embedded information is considered to be leaked to Eve.

\subsection{Contributions}

An overarching contribution of our work is that is it based on extensive field testing in a live indoor Wi-Fi network. We modified the firmware on 3 Raspberry Pi devices to extract the CSI data from the hardware to make it available in the driver software. We then conducted multiple experiments in a variety of configurations and recorded the data in a corpus of size $\approx$ 60MB. We plan to make both the data and code available to the community at \url{https://github.com/tomer-avrahami/shake-on-it} .

We first show that if Alice and Bob are stationary, then the CSI data is of low diversity. However, if endpoint Bob is shaken, there is enough diversity in the CSI values to serve as a source of true randomness for key generation. 

Prior research indicated that the CSI is highly correlated among neighboring sub-carriers, which we verified experimentally. To address this challenge we chose to employ the CSI only from a small set of non-adjacent sub-carriers. We conducted an exhaustive search to identify the best number of sub-carriers, $k$, and which $k$ sub-carriers to use, in order to minimize the correlation among them.

Next we investigated how to set the parameter $q$, the number of quantization bits to use for a given CSI value, and its effect on the bit mismatch rate (BMR). To improve the BMR, we use the fact that adjacent sub-carriers are correlated to our advantage: we use a majority rule over the values extracted from a neighborhood of $2m+1$ sub-carriers around the selected sub-carrier, and investigated the effect of the parameter $m$ on the BMR.

With this framework we demonstrate that Alice and Bob observe a BMR of 5-15\% while Eve observes a BMR of around 50\%, even if she is placed within 10cm of Alice.

Based on the above-mentioned observations, our next major contribution is the design of our key extraction and agreement method: Shaking device Bob while transmitting and recording $N$ CSI-carrying packets, extracting the bitstreams from them, using the QKD Cascade algorithm \cite{brassard1994secret} to reconcile the bitstreams, and computing a PSK as a cryptographic hash of the reconciled bitstream.  

We evaluate the security of our scheme as follows. While the Cascade algorithm leaks information, it precisely quantifies the number of bits it leaks. Thus the number of Cascade-leaked bits is subtracted from the number of information bits collected as a basic bound $B_0$ on the number of secure information bits. Further, We observe that the extracted bits are not uniformly distributed. Therefore we use the cryptography-oriented definition of min-entropy 
and use it to conservatively reduce $B_0$, to get a bound $B$ on the number of remaining secure information bits embedded in the bitstream. 
The value $B$, normalized by the number of packets $N$, is the Secured Bit Generation Rate, measured in bits per packet.

Finally we evaluate the performance of the system at various distances, and the time it takes to generate a secure shared key.

Putting it all together, our system exhibits an SBGR of $\approx$1.2 to 1.6 bits per packet, at distances of 1m--9m in an indoor environment, and can hence generate a secure shared 128-bit key within 20sec of device shaking.

\section{Background}

%
\subsection{Orthogonal Frequency Division Multiplexing (OFDM) in Wi-Fi}
\label{sec:Wifi}
%
%

At the OFDM transmitter,  the incoming data stream is split into
multiple narrow and orthogonally overlapped sub-carriers.
The data on each sub-carrier is then modulated, i.e. converted,
to the time domain by using inverse Fast Fourier Transform (IFFT).
The time-domain signal is then up-converted to radio frequency and
transmitted through the channel.
At the receiver, after frequency down-conversion, the signal is converted back to the frequency domain via FFT.

Orthogonality, viewed in the frequency domain, is achieved in OFDM
by choosing the symbol length and the frequency separation between
the sub-carriers such that the peak of each sub-carrier falls on the nulls of the others. Number of sub-carriers depend on the selected PHY protocol. For example,  IEEE 802.11n, 2.4GHz band, High-Throughput, 20MHz  uses 64 sub-carriers (pilots and guard-bands included) with spacing of \sloppy 312.5kHz, leaving 56  usable sub-carriers. In IEEE 802.11n, 5GHz band, High-Throughput 40MHz frequency range uses total of 128 sub-carriers, while 114 are usable. See Table~\ref{tab:WiFi bandwidth} for more details.

In our experiments we used 20MHz channel bandwidth over 2.4GHZ band or 40MHZ channel bandwidth over 5G band. See Table~\ref{tab:setups} for experiments setup description.

\begin{table}[t]
\begin{center}
\begin{tabular}{|c|c|c|c|c|}
  \hline
  PHY protocol & Bandwidth & FFT size & Data & Pilots \\ 
  \hline
  802.11n,802.11ac & 20 Mhz & 64 & 52 & 4\\ 
  \hline
  802.11n,802.11ac & 40 Mhz & 128 & 108 & 6\\ 
  \hline
  802.11ac & 80 Mhz & 256 & 234 & 8\\ 
  \hline
  802.11ac & 160 Mhz & 512 & 468 & 16\\ 
  \hline
\end{tabular}
\end{center}
\caption{802.11n/802.11ac WiFi sub-carrier for supported phy protocols}
\label{tab:WiFi bandwidth}
\end{table}

%
\subsection{Channel State Information (CSI)}
%
%
A characteristic of Wi-Fi is the
usage of Orthogonal Frequency Division Multiplexing (OFDM) as a
bandwidth-efficient technology for supporting high data rates.

For its proper operation, OFDM technology requires the calculation
of Channel State Information (CSI) for each sub-carrier on each antenna. The
CSI describes what the transmitted signal has undergone while
passing through the channel and reveals the combined effect due to
scattering, fading, and power decay.
An OFDM system viewed in the frequency domain can be modeled by
\begin{equation}
y=Hx+n,
\label{eq_y}
\end{equation}
where $y$ and $x$ are the received and transmitted vectors
respectively, $H$ is the channel matrix and $n$ is an additive white Gaussian noise (AWGN) vector.

To successfully detect the message $x$ from the received signal $y$,
distorted by fading and noise, OFDM receivers first need to estimate the channel. This is achieved by transmitting predetermined symbols, a.k.a.\ preamble, or pilots. Thus, the CSI, given for all sub-carriers in the
form of the matrix ($\hat H$),
can be estimated in view of Equation (\ref{eq_y}).

Since OFDM reception requires accurate estimation of the CSI, it is safe to assume that this information is also available for other uses. The Wi-Fi device drivers for the Raspberry Pi we experimented with extract the CSI data computed by the hardware and provide it to the upper layers of software---enabling its use for shared-key generation.

\subsection{Channel coherence time and reciprocity}
\label{sec:channel reciprocity}

The channel coherence time $T_c$ is commonly defined as
the time in which the channel can be considered constant \cite{tse2005fundamentals}---which for our purposes implies that the channel is reciprocal. We conservatively assume that this is
the time it takes a moving antenna to traverse 0.25$\lambda$. We can over-estimate the momentary velocity of a hand shaking a mobile device at 3m/s, so for Wi-Fi at 2.4GHz ($\lambda=12.5$cm) we get $T_c\approx 10.4$ms, and for 5GHz ($\lambda=6$cm) we get $T_c\approx 5$ms. So if the round-trip time over Bob and Alice's CSI-carrying messages is below 5ms then both parties observe a reciprocal channel and measure highly correlated CSI.

\subsection{Min-entropy}
\label{sec:min-entropy}

A crucial requirement of using CSI for cryptographic key generation is that CSI data, viewed as a bitstream, appears as resulting from a true random bit generator (TRBG). Therefore there is a need to estimate the amount of entropy in the extracted bit stream. Following~\cite{barak2003true,SP80090B} we argue that in an adversarial setting, the correct measure of entropy is not the classical Shannon entropy, but rather the {\it min-entropy}. 

Min-entropy is a measure of the worst-case randomness in a random source, and using it is more conservative and more secure. It is calculated by using only the item with the highest probability of any possible outcome.
 The min-entropy of an independent discrete random variable $X$ that takes values from the set \(A=\{x_1,x_2,\ldots,x_k\}\) with probability \(Pr(X=x_i) = p_i\) for $i=1,\ldots,k$ is defined as:
 \begin{equation}
    H= \min_i(-\log_2p_i)=-\log_2(\max_i(p_i))
 \end{equation}
If $X$ has min-entropy $H$, then the probability of observing any particular value for $X$ is no greater than \(2^{-H}\). The maximum possible value for the min-entropy of a random variable with $k$ distinct values is \(log_2k\), which is attained when the random variable has a uniform probability distribution, i.e., \(p_1 = p_2 = \ldots = p_k =1/k\).

\subsection{Cascade}
\label{sec:Cascade}
In the context of quantum key distribution (QKD) there is an analogous situation to our setup. Alice and Bob have two channels: a private quantum channel that introduces errors, and a separate public channel that is error free. Alice and Bob each record a raw bitstream from the private channel, and then reconcile the keys by exchanging extra information on the public channel. The information transmitted on public channel is assumed available to the eavesdropper (Eve). So, the amount of information revealed must be considered in the security analysis of the protocol.

The first key reconciliation algorithm for QKD (BBBSS) was presented in~\cite{bennett1992experimental}, and subsequently improved by~\cite{brassard1994secret} and called the Cascade algorithm.
This protocol consists of a sequence of rounds. Between every two rounds the bitsreams are permuted randomly.
In each round Alice and Bob divide their bitstreams into blocks and exchange the parity of each block. When they find a block with different parity, they perform binary search on this block to eliminate one error within it, and then re-do the error correction of previous rounds using the newly corrected bit. Thus a single corrected bit in round $i$ can lead to the correction of several other bits (in other blocks). Note that during a single round the parties process multiple blocks in parallel and the parity values of all the blocks may be packed into a single message on the public channel.
Cascade code in this work is based on a Cascade-Python implementation forked from \cite{cascade:python}.

\section{System setup and basic measurements}

\subsection{System setup}
Alice and Bob  exchange packets wirelessly, while Eve ipassively eavesdrops on the communication. Alice plays the part of the access point so it is fixed in place, while Bob can be shaken. 
%
We found that the most accurate data Eve can hope to eavesdrop is obtained when she statically positions herself a few centimeters away from Alice. 

\subsubsection{Hardware}
Our setup consists of four components:
\begin{itemize}
    \item Alice, Bob, and Eve are Raspberry pi 4 devices
    \item A controller, which is a Python3 program running on a Dell laptop running Windows 10.
\end{itemize}
All the devices are connected to the same wireless LAN, which serves as the public channel among devices and also is used to control the experiments.
We selected the
Raspberry PI since its Broadcom-made Wi-Fi chip supports
a patched driver developed by Nexmon \cite{nexmon:project}, which allows exporting the CSI to the software layers.

\subsubsection{Firmware}
Based on the Nexmon project \cite{nexmon:project}, we loaded our own firmware version to the Broadcom Wi-Fi chip. 
This version enables:
\begin{itemize}
\item Monitor mode: allows sniffing all the Wi-Fi packets.
\item Injection from host: allows injecting custom Wi-Fi packets triggered by the host operating system.
\item Injection from driver: allows injecting custom Wi-Fi packet triggered by logic within the Wi-Fi chip firmware. 
\item CSI access: allows reading CSI values from the  hardware registers and sending it to the host software.
\item Wi-Fi channel and bandwidth control.
\end{itemize}

\begin{figure}[t]
  \centering
  \includegraphics[width=\linewidth]{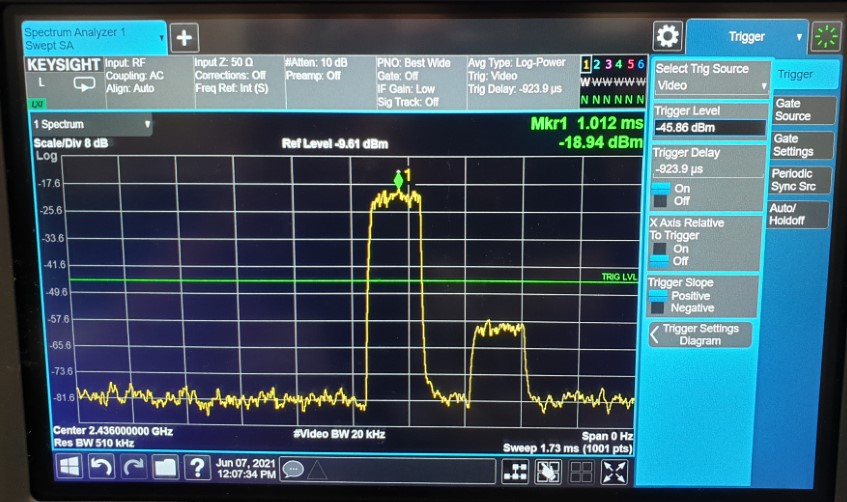}
  \caption{Alice (left) and Bob's (right) packets observed on spectrum analyzer. Time between Alice's packet to Bob's packet is approximately $150\mu$s. The round-trip-time is approximately $480\mu$s.}
  \label{fig:spectrum}
\end{figure}

All three devices use the same network driver firmware code, to which we added a ``quick-reply'' feature. If this feature is enabled, the device replies with a response packet whenever a new legitimate packet arrives.
This feature was enabled only in Bob's firmware.
As Bob's response is sent directly from the network driver firmware, the time duration between Alice's packet and Bob's packet is kept well below 5ms, which is essential for maintaining channel reciprocity (recall Section~\ref{sec:channel reciprocity}).
As can be seen in Figure~\ref{fig:spectrum}, 
With this feature we measured the time from the end of Alice's packet to Bob's response to be $\approx150\mu$s, and each packet transmission duration is $164\mu$s yielding an end-to-end round trip time of $\approx 480\mu$s.

In our experiments we used 139-byte QOS packets whose payload was constant.
Note that with CSI access, the firmware driver on the receiver side overwrites the buffer containing the packet payload with the CSI information before delivering it to the calling software. Thus the payload in the packets used to exchange CSI is unavailable to the receiver.

\subsection{The effect of shaking}

One of the goals of the shared-key generation is that the key should be based on a true random bit generator (TRBG).

\begin{figure}[h]
  \centering
  \includegraphics[width=\linewidth]{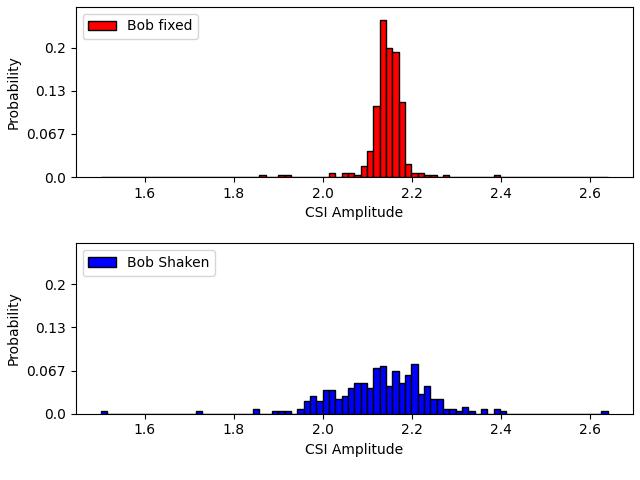}
  \caption{The probability density functions of the CSI amplitude for a static channel (top) and a channel in which Bob is moving (bottom), using 2.4 GHz at a distance of 3m.}
  \label{fig:static-and-shaking}
\end{figure}

Several tests we conducted in static environments have validated, as found in previous works ~\cite{7557048},
that 
the RF channel parameters remain near-constant 
for a long time: Practically a fixed CSI value is measured, perturbed by non controllable environmental noise and interference---see Figure~\ref{fig:static-and-shaking} (top). 

However, when Bob's position with respect to Alice changes rapidly enough by manually shaking it, much more diversity is included in the CSI data---see Figure~\ref{fig:static-and-shaking} (bottom).

\begin{figure}[t]
  \centering
  \includegraphics[width=\linewidth]{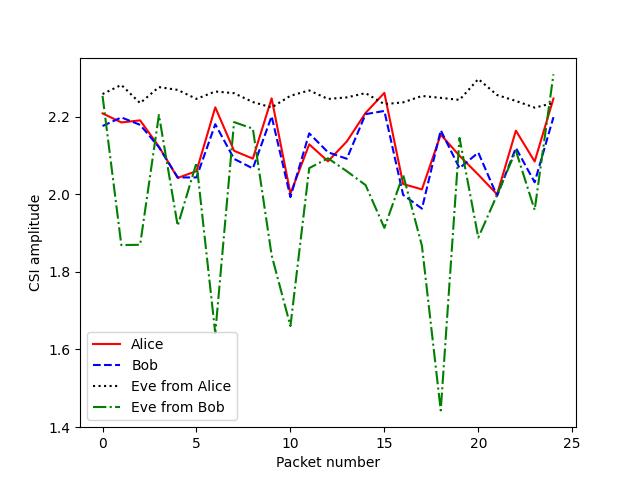}
  \caption{Calibrated CSI amplitudes measured by Alice and Bob on sub-carrier -7 on channel 6, while Bob is in motion (Zoom in).} 
  \label{fig:one sc csi}
\end{figure}

\subsection{Can Eve eavesdrop?}

An important goal of the system is to keep the key private from Eve. In several experiments we demonstrated that as long as Eve is outside the spatial correlation distance $D_c$, the CSI data she can observe is decorrelated from the observations of Alice and Bob.

Figure~\ref{fig:one sc csi} shows the CSI amplitude of a single sub-carrier over time as measured by Alice, Bob, and twice by Eve: once from Alice and once from Bob. It is evident from the figure that Alice and Bob measure similar CSI amplitudes, while Eve's measurements are near constant toward Alice and very divergent from the Alice-Bob channel towards Bob. Note also that Bob's movement induces a large variance in the value.

\begin{figure}[t]
  \centering
  \includegraphics[width=\linewidth]{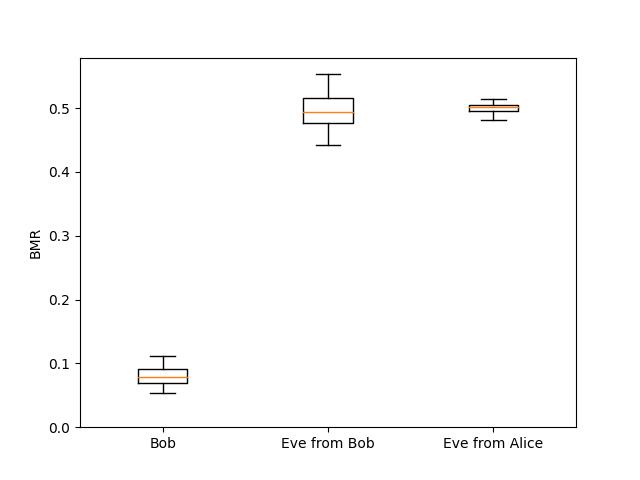}
  \caption{Bit mismatch rate (BMR) comparison. Data is taken from 33 experiments at different distances. Alice has the correct key, Bob extracted a key with BMR of 5\%-15\%, while Eve got BMR of $\approx 50\%$.}
  \label{fig:eve-bmr}
\end{figure}

Once we apply the quantization on the selected sub-carriers (to be described below) to extract a bitstream, we can calculate the bit mismatch rate (BMR) observed by Bob and compare it to that observed by Eve. Our measurements show that while Bob observes a BMR of 5\%-15\%, Eve observes a BMR of $\approx$50\% as can be seen in  Figure~\ref{fig:eve-bmr}.

\begin{figure}[t]
  \centering
  \includegraphics[width=\linewidth]{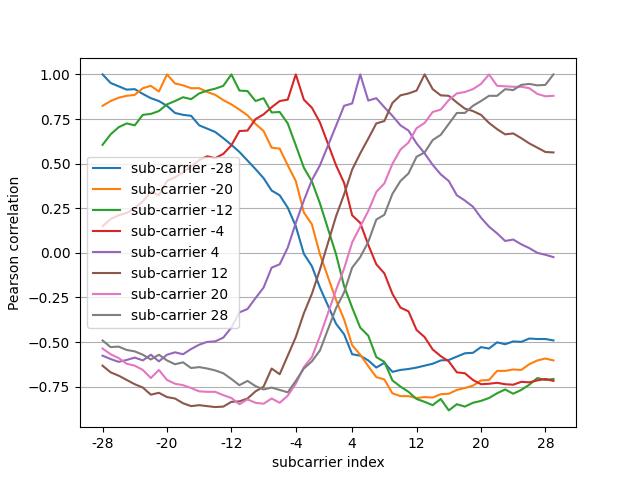}
  \caption{Cross correlation between sub-carriers. Each curve presents the correlation of one sub-carrier with all the others. For example, Taken on WiFi channel 6, 20MHz bandwidth.}
  \label{fig:sub carrier cross correlation lines}
  \vspace{-0.5cm}
\end{figure}

\subsection{Sub-carriers selection}
\label{sec:Correlation between sub-carriers}
As already known from previous works, the CSI values of neighboring sub-carriers are correlated. To evaluate this phenomenon in our environment, Figure~\ref{fig:sub carrier cross correlation lines} shows that indeed the correlation between adjacent sub-carriers is strong, and correlation strength is oscillating as function of sub-carriers index. This empirically-generated figure is well aligned with the analytical graphs presented in \cite{TDS}.  

\begin{figure}[t]
\vspace{-0.4cm}
  \centering
  \includegraphics[width=\linewidth]{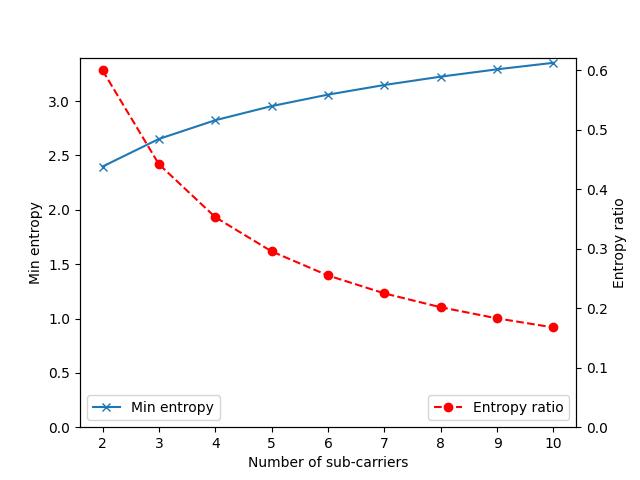}
  \caption{Min-entropy and Entropy ratio as functions of the number of sub-carriers ($k$) used for CSI data, with $q=2$ quantization bits per sub-carrier.}
  \label{fig:num_of_sc_vs_min_entropy}
\end{figure}

Therefore, unlike implementations such as \cite{6567117}, we do not 
extract bits from all available sub-carriers---doing so would produce an unfounded sense of security since the extra bits do not add true randomness to the system. 
Instead, we conducted an exhaustive search for relatively small subsets of non-neighboring sub-carriers which provide high randomness, using min-entropy (recall Section \ref{sec:min-entropy}) as our randomness measurement score. As can be seen in Figure~\ref{fig:num_of_sc_vs_min_entropy}, as we increase $k$, the number of sub-carriers, we do get more entropy---however the ratio between the added entropy and added bits decreases sharply. As a concrete example, when $k=4$ we found that selecting sub-carriers -25, -7, 11, 27 yields 2.8 bits of entropy.

However, we also use the fact that neighboring sub-carriers are correlated to reduce errors: As we shall see in Section~\ref{sec:Majority rule} we use a majority-rule decision among $2m+1$ neighboring sub-carriers.

\section{The Method}
\label{sec:Method}
\subsection{Protocol parameters}
\label{sec:Protocol parameters}
When Alice and Bob initialize, the following parameters should be defined:
\begin{itemize}
    \item $N$ - number of packets.
    \item $k$ - number of sub-carriers in use.
    \item $q$ - number of quantization bits
    \item $m$ - majority margin, number of sub-carriers to use for majority rule from each side of the main sub-carrier
\end{itemize}
\subsection{SBGR---secured bit generation rate}
\label{sec:SBGR}
We argue that the raw bit generation rate, based simply on the number of extracted bits per packet, is not an effective performance indicator since it doesn't take security into account. Instead we define a new key performance indicator we denote by SBGR: the Secure Bit Generation Rate.

The SBGR starts with the number of raw extracted bits, but takes two aspects into consideration:
\begin{enumerate}
    \item During key reconciliation using Cascade (recall Section~\ref{sec:Cascade}) information bits are leaked, and Cascade returns their exact number. Thus we deduct the number of leaked bits from the number of raw bits and normalize by the number of packets $N$.
    \item Despite the selection of only $k$ sub-carriers, some correlation remains in the extracted bits. To take this effect into account we multiply by the ratio between minimum entropy and maximum possible entropy value.
\end{enumerate}
Putting it all together, SBGR is calculated by
\begin{equation}
    SBGR = \frac{rawBits-leakedBits}{N}\times\frac{minEntropy}{k\cdot q}
\end{equation}
Where $N$,$k$,$q$ are as defined in Section~\ref{sec:Protocol parameters}, $rawBits$ is the length of the raw bitstream, $leakedBits$ is number of bits leaked during key reconciliation process, and $minEntropy$ is the min-entropy value (out of the maximum, which is $k\cdot q$).
\subsection{CSI data collection}
\subsubsection{Packets sync}
Alice and Bob need to make sure they are using the same packets to build the security key.
Alice can know if a specific packet exchange succeeded according to Bob response packet. Bob, on the other hand, needs Alice's approval to make sure the packet exchange succeeded. We achieve this by a simple req-ack with retrial mechanism:
\begin{enumerate}
    \item Alice sends packet $n$. The packet's data includes the value $n$.
    \item Bob gets the packet, saves it, and responds with an ack packet which also includes the value $n$.
    \item If Alice got Bob's ack packet she continuous to packet \(n+1\), otherwise she resends packet $n$.
    \item In case of a retrial, Bob overwrites packet $n$'s CSI data and resends an ack for packet $n$.
\end{enumerate}

In our experiments, a separate device was used as a synchronization controller to make sure Eve wirelessly receives all the possible data as well. Thus a packet exchange is successful only if Bob received Alice's packet, Alice received Bob's packet, and Eve received both Alice's and Bob's packets.

\subsubsection{CSI Calibration}
 The CSI data measured by the devices cannot be directly used since
the Automatic Gain Control (AGC) scheme at the receiver modifies the amplitude of the original CSI: The CSI value reported by the device, is multiplied by an unknown
factor whose value changes as the user moves. 
However, since the RSS is reported before AGC occurs, we can reconstruct the CSI by re-scaling it using the received RSS value. Following Gao et al.~\cite{9187854}, we multiply the extracted CSI of all sub-carriers by:
\[
\sqrt{\frac{\mbox{\it RSS}}{
\sum_{n=1}^{numSC}\mbox{\it CSI}^2 } }
\]

\subsection{Secured key extraction}
After collecting $N$ packets, Alice and Bob can initiate the proposed key extraction protocol. The protocol has four steps:
(i) Quantization (ii) Majority rule (iii)
Key reconciliation (Cascade) and (iv)
Shared secret key extraction.

\subsubsection{Quantization}

The quantization process is carried out for each sub-carrier separately.
We use equally-sized percentiles to define quantization bins. Assuming $q$ quantization bits we set $2^q-1$ quantization levels associated with $2^q$ bins. The quantization levels \(QL_i\) are calculated as follows: 

\[
QL_i = Percentile(i*100/(2^q))
\]
where $i$ denotes the quantization level index \(i = [1..2^q-1]\) and \(Percentile(x)\) returns the value where $x$ percentage of the samples are below this value.

We encode each CSI measurement that is in quantization bin $i$ by the $q$-bit symbol representing $i$ using a Gray code.

One advantage of using percentiles for quantization is that each symbol has the same frequency, supporting a uniform distribution of bit values. 
Note that each device sets its own quantization levels using its own data, eliminating the need to calibrate individual device-dependent thresholds.

Bit mismatches are typically caused by Alice and Bob's measurements falling into different, yet adjacent, bins: The Gray encoding ensures that such mismatches only flip one out of the $q$ bits, which helps reduce the bit mismatch rate.

\begin{figure}[t]
  \centering
  \includegraphics[width=\linewidth]{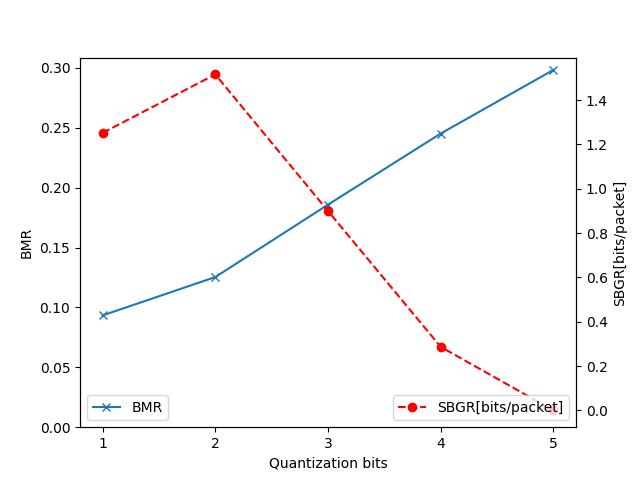}
  \caption{Secure bit generation rate (SBGR) and bit mismatch rate (BMR) as functions of the number of quantization bits.}
  \label{fig:quan_bits_vs_real_bits_per_packet_and_ber}
  \vspace{-0.4cm}
\end{figure}

As can be seen at Figure \ref{fig:quan_bits_vs_real_bits_per_packet_and_ber}, $q=2$ quantization bits is the optimal selection: a larger $q$ increase the BMR which causes the SBGR to drop (due to additional leakage in the reconciliation). Choosing $q=1$  decreases the amount of data.

\subsubsection{Majority rule}
\label{sec:Majority rule}
As we discussed in Section~\ref{sec:Correlation between sub-carriers}, adjacent sub-carriers are strongly correlated so we selected only $k$ non-neighboring sub-carriers, which we call the \emph{main} sub-carriers. However, we employ this correlation to reduce errors by making a majority rule decision among $(2m+1)$ sub-carriers centered around each main sub-carrier. Let \(l\) be the index of a main sub-carrier, then the extracted symbol \(y\) is chosen by:
\begin{equation}
y=argmax_a(sum_j(sc_j==a))
\end{equation}
Where $j$ gets $2m+1$ values in range \([l-m,\ldots,l+m]\), and \(sc_j\) is the symbol extracted from the \(j^{th}\) sub-carrier. Ties are broken arbitrarily.

Note that with $k$ main sub-carriers, if the minimal distance between them is $d$ sub-carriers then we should use at most $m<d/2$ in the majority rule to avoid using the same sub-carrier in the majority rule calculation of two main sub-carriers. E.g., with $k=4$ and the main sub-carriers from the example in Section~\ref{sec:Correlation between sub-carriers} we have $d=16$ so we need $m<8$.

\begin{figure}[t]
  \centering
  \includegraphics[width=\linewidth]{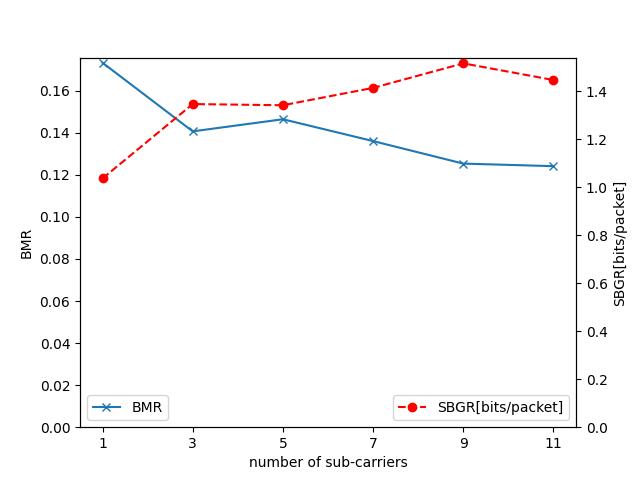}
  \caption{The secured bit generation rate (SBGR) and bit mismatch rate (BMR) as functions of the number \( (2m+1) \) of sub-carriers used in the majority rule.}
  \label{fig:majority_call_vs_rbpp_and_ber}
  \vspace{-0.5cm}
\end{figure}

As evident from Figure~\ref{fig:majority_call_vs_rbpp_and_ber}, using the majority rule
indeed reduces the bit mismatch rate and thus improves the SBGR, with the best results achieved with 9 sub-carriers ($m=4$). 

\subsubsection{Key reconciliation}
The raw bitstream is a concatenation of all the quantized bits collected as described above. Both Alice and Bob generate their own bitstreams. Due to the channel reciprocity their bitstreams are very similar, but not identical. Bit mismatches are due to quantization errors, interference and noise, and also motion: e.g., Bob sent his response from outside of the coherence distance.

In order to reconcile all the mismatches, Alice and Bob agree that Alice is the one holding the ``correct bitstream'', and Bob corrects his bitstream using the Cascade algorithm~\cite{brassard1994secret}.

In order to reduce the number of Cascade rounds we modified the Cascade implementation of~\cite{cascade:python} as follows:

\begin{enumerate}
\renewcommand{\theenumi}{\alph{enumi}}
    \item Early termination: Upon initiating the process, Alice sends Bob a hash of her (``correct'') bitstream. 
    After each Cascade round, Bob compares the hash of his reconciled bitstream equals with the one Alice sent. If the two are identical Bob sends a success indication to Alice and the process stops. This feature saves both run-time and amount of leaked bits.
    \item Increasing the number of rounds:  The original Cascade algorithm~\cite{10.1007/3-540-48285-7_35} uses 4 rounds. Following Yan et al.~\cite{4667215} we increased the number of rounds to 10, but the additional rounds use an initial block size of half of the bitstream length. In most cases this change is immaterial since~4 rounds suffice to reach full reconciliation, but in rare cases this helps ensure that Cascade converges. 
    \item After the final round, if Bob observes that the bitstream hashes are not identical then he sends a ``failure'' indication to Alice.
\end{enumerate}
\subsubsection{Shared secret Key extraction}
At the end of a successful key reconciliation process, each device holds a bitstream whose length is 
$L=(N\cdot k\cdot q)$. As discussed in Section~\ref{sec:SBGR}, the number of secure bits embedded in the bitstream after the Cascade leakage, and accounting for the min-entropy, is only $L^* = N\cdot SBGR$, which 
is significantly less than $L$. So neither the bitstream itself, nor pieces of it, should be used directly as cryptographic keys. 
Instead, 
we invoke a strong hash function such as the SHA-256 on the whole bitstream to compress all the available randomness into a 256-bit shared key: as long as $L^* > 256$ we can claim that the shared key indeed has 256 cryptographically secure bits.

\begin{table}[t]
\begin{center}
\begin{tabular}{|c|c|c|}
  \hline
  Parameter & Setup A & Setup B \\ 
  \hline
  Band & 2.4GHz & 5GHz \\ 
  \hline
  WiFi channel & 6 & 36 \\ 
  \hline
  WiFi frequency & 2437MHz & 5180MHz \\ 
  \hline
  WiFi bandwidth & 20MHz & 40MHz \\ 
  \hline
  Available sub-carriers & 56 & 114 \\
  \hline
    \#main sub-carriers ($k$) & 4 & 8 \\ 
  \hline
  \#majority rule sub-carriers ($m$) & 4 & 4 \\
  \hline
  Quantization bits ($q$)& 2 & 2 \\ 
  \hline
  Tested range & 0.5m-9m & 0.5m-9m \\
  \hline
\end{tabular}
\end{center}
\caption{Experiments' chosen setup and parameters.}
\label{tab:setups}
\vspace{-0.8cm}
\end{table}

\section{Results and discussion}
All the experiments were carried out in a typical indoor environment---a residential apartment in a crowded city. The  Experiments were carried out in two different Wi-Fi setups, as detailed in Table~\ref{tab:setups}.

\subsection{Operational range}
\begin{figure}[t]
  \centering
  \includegraphics[width=\linewidth]{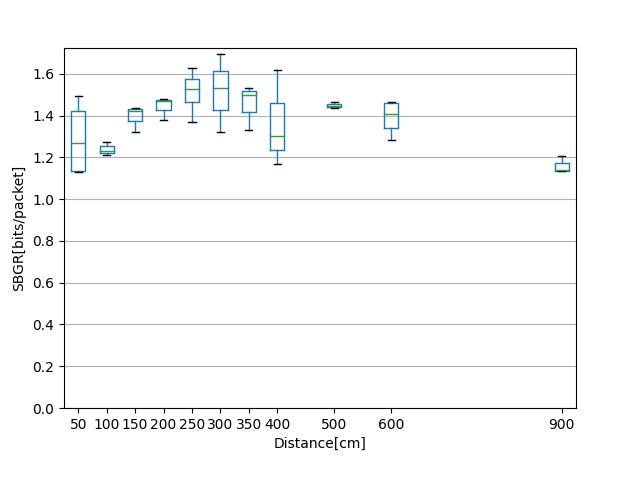}
  \caption{SBGR as function of the distance between Alice and Bob.}
  \label{fig:SBGR_vs_bob_y}
\end{figure}

We tested our setup in a variety of realistic indoor ranges and Figure~\ref{fig:SBGR_vs_bob_y} shows the achieved SBGR. As can be seen, the optimal working range is 2--4m, but the system still performs well even at a distance of 9 meters.

\subsection{Time to key}
\begin{figure}[h]
\vspace{-0.55cm}
  \centering
  \includegraphics[width=\linewidth]{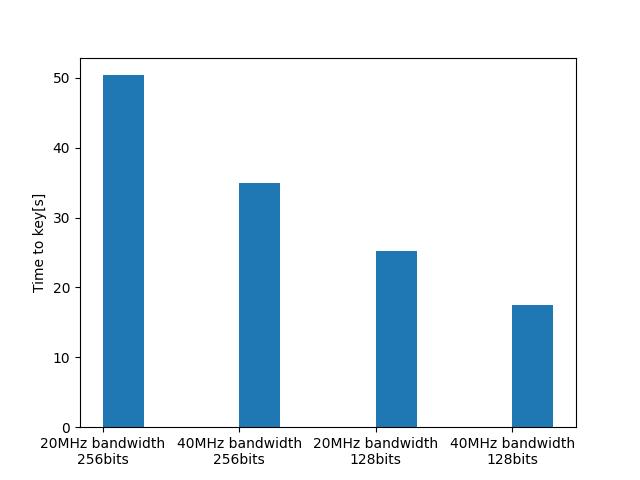}
  \caption{The average measured time to generate a 128 and 256 bits keys at 3 meters, in two scenarios.}
  \label{fig:ttk}
\end{figure}

Fast shared key generation is a critical performance parameter of a system when there is human involvement needing to shake the device.
Figure \ref{fig:ttk} shows the average time to generate 128- and 256-bit keys at 3 meters, for the two different setups as described in Table~\ref{tab:setups}.
The time to key consists of several contributors:
\begin{itemize}
    \item Time between adjacent cycles: $\approx 280ms$ in our setup. 
    \item The number of Cascade requests that Bob sends, which depends on the number of bits to reconcile.
    \item The round trip-time for a Cascade packet: $\approx$200ms.

\end{itemize}


Based on some initial experiments we believe that in ``production code'', the round-trip time can be reduced by at least 60\%.

Figure~\ref{fig:ttk} shows that the times needed to generate 128 and 256 bits, in the 2.4GHz band (20MHz bandwidth) and 5GHz (40MHz bandwidth) are 20-50s, even in our non-optimized implementation.

\section{Related Work}
\label{sec:Related work}

The topic of physical layer secret key generation has been investigated by the research community, using various techniques and focusing on diverse technologies. The surveys of \cite{zdmw16,hamamreh19} provide a taxonomy of approaches.

Several works analyzed and simulated the use of CSI for generating a secure key. The closest ones to our approach are CGC~\cite{6567117} and KEEP~\cite{xi2014keep}, so we also include Table~\ref{tab:comparison} in which we compare our results with theirs. 


The CGC work of Liu et al.~\cite{6567117} focused on reducing CSI mismatch errors by removing the non-reciprocity component learned from a small number of probe packets. They reported a very high bit generation rate by using CSI data from \emph{all} sub-carriers to extract 60 bits per packet. However, as \cite{TDS} analyzed, and our work demonstrated in practice (recall Section~\ref{sec:Correlation between sub-carriers}), subcarrier data is highly correlated, especially in a static environment, so such high-rate bits cannot be assumed to be secure: hence the indication of ``?'' in Table~\ref{tab:comparison}. Nevertheless CGC made some valuable observations on reducing the mismatch rate: it would be interesting to add such a mechanism to our work. This is deferred to future work.

The KEEP work of Xi et al.~\cite{xi2014keep}  addresses the same scenario as our work, of device-to-device shared secret key generation. They addressed the correlation between adjacent sub-carriers by using all the sub-carriers to generate a single bit of data, which eventually resulted in a slow bit generation rate of 0.5 bits per packet. More importantly, though, the KEEP solution severely over-estimates the strength of the resulting keys. KEEP creates a long key by concatenating multiple short CSI-derived blocks, of 16 or 32 bits each. Since the protocol exposes the hash of each of the blocks, the long key's cryptographic strength is in fact determined by the size of the block, since the blocks can be cracked in sequence. So, for example, a key of 256 bits that is generated from 8 32-bits blocks only has a strength of 32 bits.


The TDS system of Wei et al.~\cite{TDS} also relies on CSI and is based on the coherence distance principle, which states that two devices located less than ~$0.25\lambda$ from each other (3.1cm at 2.4GHz band), will experience the same channel. However, unlike in our scenario, their solution relies on a third device generating packets, and is only feasible when both Alice and Bob are practically touching each other so their antennas are within the coherence distance.

The work of Zhang et al.~\cite{7448884} simulated and analyzed a key generation process using one sub-carrier to achieve a shared key between Alice and Bob in a slow-fading channel. 
They showed that in order to achieve temporal de-correlation between the CSI of successive packets one must have $\approx$200ms delay between them. This result does not contradict our belief that our system can operate with $\approx$100ms cycle time: since Bob is shaken the channel is rapidly changing.


Recently more portable CSI extraction tools have become available, such as the ESP CSI tool~\cite{Hern2006:Lightweight}, or the Nexmon CSI tool~\cite{nexmon:project}---which is the one we also used. 
Yuan et al.~\cite{9880811} focused on CSI robustness of the esp32 CSI tool and used it to analyze reciprocity on different scenarios, while 
Cao et al.~\cite{10012875} focused on its full system firmware development. 

Li et al.~\cite{9625573} shows channel reciprocity in different static environments using the Nexmon firmware~\cite{nexmon:project}, however they did not analyze a full key-generation system performance and didn't relate to the key strength or the performance. Their results strengthen our work and match our observations on channel reciprocity. 

The idea of shaking WiFi devices for key agreement already appears in the ``Shake them up!'' work of Castelluccia et al.~\cite{10.1145/1067170.1067177}. However their scenario is different than ours: they assume the legitimate user can hold both devices simultaneously while the adversary is several meters away. The devices transmit messages with spoofed source MAC addresses, and the claim of security is based on the (challenging) premise that the adversary cannot identify which of the two devices is transmitting.
Several other works used human hand shaking to generate randomness to securely pair two mobile devices. For example, ~\cite{10.5555/1771592.1771610} and ~\cite{4796201}, used it while extracting the random data from accelerator sensors. 
Besides CSI, some authors used the basic RSSI (receiver signal state indicator) to extract achieve key agreement~\cite{6226870},\cite{9000831}.

CSI data was also used for fingerprinting-based self-localization employing the Wi-Fi infrastructure 
\cite{14:MonoPhy,2:PinLoc,4:FIFS}. 
The basic idea behind these methods is to
hold an on-site training phase in which a CSI ``fingerprint'' is collected for each geographic location. The localization phase includes measurement of the CSI and finding the best match to the fingerprints database. In \cite{6:FILA} and \cite{6:CSI} a method for CSI-based indoor
range estimation is suggested.
CSI was also used for direction-finding of rogue Wi-Fi hotspots~\cite{taw15}.

\begin{table}[t]

\begin{center}
\scalebox{0.8}{
\begin{tabular}{|c|c|c|c|}
  \hline
  Parameter & CGC & KEEP & Shake on it \\ 
  \hline
  Band & 2.4GHz & 2.4GHz & 2.4GHz / 5GHz \\ 
  \hline
  Raw BGR [bits/packet]  & 60 & 0.5 & 8 / 16 \\ 
  \hline
  Secure Bit Generation rate & ? & ? & 1.5 / 2.5 \\ 
  \hline
   Key Strength & ? & 32bits & 256bits  \\ 
  \hline
  Tested range & 2-3m & 2m & 0.5m-9m  \\ 
  \hline
\end{tabular}
}
\end{center}
\caption{Comparison with other works, at 2.4GHz}
\label{tab:comparison}
\vspace{-0.8cm}
\end{table}


\section{Conclusions}
In this work we introduced and implemented a practical end-to-end scheme for generating a symetric key between two WiFi devices based on (existing) CSI measurements. We demonstrated that the randomness in CSI values obtained in a static scenario is insufficient for secure key generation. 
In order to significantly alleviate this limitation, we proposed shaking the endpoint Bob. Thus, sufficient variation in the CSI values is achieved and a source of true random (bit) information is obtained.
%

For analyzing the cryptographic strength of the resulting key we defined the {\it Secure Bit Generation Rate}, SBGR. This measure takes into account the total number of bits collected, number of packets, amount of leaked bits due to Cascade reconciliation, and the cryptographically-accepted metric of min-entropy.
We conducted over 60 end-to-end experiments under different practical scenarios and verified that the proposed scheme performs well for a variety of distances (between endpoints), under different physical layer (PHY) settings. All-in-all we demonstrate that a strong symmetric key can be obtained in a reasonable time.
%



\bibliographystyle{IEEEtran}
\bibliography{lets_shake_on_it}

\begin{thebibliography}{10}
\providecommand{\url}[1]{#1}
\csname url@samestyle\endcsname
\providecommand{\newblock}{\relax}
\providecommand{\bibinfo}[2]{#2}
\providecommand{\BIBentrySTDinterwordspacing}{\spaceskip=0pt\relax}
\providecommand{\BIBentryALTinterwordstretchfactor}{4}
\providecommand{\BIBentryALTinterwordspacing}{\spaceskip=\fontdimen2\font plus
\BIBentryALTinterwordstretchfactor\fontdimen3\font minus
  \fontdimen4\font\relax}
\providecommand{\BIBforeignlanguage}[2]{{%
\expandafter\ifx\csname l@#1\endcsname\relax
\typeout{** WARNING: IEEEtran.bst: No hyphenation pattern has been}%
\typeout{** loaded for the language `#1'. Using the pattern for}%
\typeout{** the default language instead.}%
\else
\language=\csname l@#1\endcsname
\fi
#2}}
\providecommand{\BIBdecl}{\relax}
\BIBdecl

\bibitem{alliance2019wpa3}
{Wi-Fi Alliance}, ``{WPA3} specification,'' 2019,
  \url{https://www.wi-fi.org/file/wpa3-specification}.

\bibitem{alliance2020wps}
------, ``{Wi-Fi} protected setup protocol and usability best practices
  networks,'' 2020,
  \url{https://www.wi-fi.org/download.php?file=/sites/default/files/private/Wi-Fi_Protected_Setup_Best_Practices_v2.0.2.pdf}.

\bibitem{brassard1994secret}
G.~Brassard and L.~Salvail, ``Secret-key reconciliation by public discussion,''
  in \emph{Advances in Cryptology—EUROCRYPT’93: Workshop on the Theory and
  Application of Cryptographic Techniques Lofthus, Norway, May 23--27, 1993
  Proceedings 12}.\hskip 1em plus 0.5em minus 0.4em\relax Springer, 1994, pp.
  410--423.

\bibitem{nist-sha}
\BIBentryALTinterwordspacing
{NIST}, ``{FIPS PUB} 180–4: Secure hash standard.'' Federal Information
  Processing Standards Publication, U.S. Department of Commerce, 2015.
  [Online]. Available: \url{http://dx.doi.org/10.6028/NIST.FIPS.180-4}
\BIBentrySTDinterwordspacing

\bibitem{nist800-22-1a}
\BIBentryALTinterwordspacing
NIST, ``A statistical test suite for random and pseudorandom number generators
  for cryptographic applications,'' U.S. Department of Commerce, Washington,
  D.C., Tech. Rep. SP 800-22 Rev.\ 1a, 2010. [Online]. Available:
  \url{https://csrc.nist.gov/publications/detail/sp/800-22/rev-1a/final}
\BIBentrySTDinterwordspacing

\bibitem{tse2005fundamentals}
D.~Tse and P.~Viswanath, \emph{Fundamentals of wireless communication}.\hskip
  1em plus 0.5em minus 0.4em\relax Cambridge university press, 2005.

\bibitem{barak2003true}
B.~Barak, R.~Shaltiel, and E.~Tromer, ``True random number generators secure in
  a changing environment,'' in \emph{Proc.\ 5th Cryptographic Hardware and
  Embedded Systems-(CHES)}.\hskip 1em plus 0.5em minus 0.4em\relax Springer,
  2003, pp. 166--180.

\bibitem{SP80090B}
\BIBentryALTinterwordspacing
{NIST}, ``Recommendation for the entropy sources used for random bit
  generation,'' U.S. Department of Commerce, Washington, D.C., Tech. Rep. SP
  800-90B, Jan. 2018. [Online]. Available:
  \url{https://csrc.nist.gov/publications/detail/sp/800-90b/final}
\BIBentrySTDinterwordspacing

\bibitem{bennett1992experimental}
C.~H. Bennett, F.~Bessette, G.~Brassard, L.~Salvail, and J.~Smolin,
  ``Experimental quantum cryptography,'' \emph{Journal of cryptology}, vol.~5,
  pp. 3--28, 1992.

\bibitem{cascade:python}
B.~Rijsman, ``Cascade-python,''
  \url{https://github.com/brunorijsman/cascade-python.git}, 2020.

\bibitem{nexmon:project}
\BIBentryALTinterwordspacing
M.~Schulz, D.~Wegemer, and M.~Hollick. (2017) Nexmon: The c-based firmware
  patching framework. [Online]. Available: \url{https://nexmon.org}
\BIBentrySTDinterwordspacing

\bibitem{7557048}
J.~Zhang, R.~Woods, T.~Q. Duong, A.~Marshall, Y.~Ding, Y.~Huang, and Q.~Xu,
  ``Experimental study on key generation for physical layer security in
  wireless communications,'' \emph{IEEE Access}, vol.~4, pp. 4464--4477, 2016.

\bibitem{TDS}
\BIBentryALTinterwordspacing
X.~Wei, Q.~Chen, H.~Jinsong, Z.~Kun, Z.~Sheng, L.~Xiang-Yang, and Z.~Jizhong,
  ``Instant and robust authentication and key agreement among mobile devices,''
  \emph{In Proceedings of the 2016 ACM SIGSAC Conference on Computer and
  Communications Security}, vol.~50, no.~1, p. 616–627, Oct. 2016. [Online].
  Available: \url{https://dl.acm.org/doi/abs/10.1145/2976749.2978298}
\BIBentrySTDinterwordspacing

\bibitem{6567117}
H.~Liu, Y.~Wang, J.~Yang, and Y.~Chen, ``Fast and practical secret key
  extraction by exploiting channel response,'' in \emph{2013 Proceedings IEEE
  INFOCOM}, 2013, pp. 3048--3056.

\bibitem{9187854}
Z.~Gao, Y.~Gao, S.~Wang, D.~Li, and Y.~Xu, ``Crisloc: Reconstructable csi
  fingerprinting for indoor smartphone localization,'' \emph{IEEE Internet of
  Things Journal}, vol.~8, no.~5, pp. 3422--3437, 2021.

\bibitem{10.1007/3-540-48285-7_35}
G.~Brassard and L.~Salvail, ``Secret-key reconciliation by public discussion,''
  in \emph{Advances in Cryptology --- EUROCRYPT '93}, T.~Helleseth, Ed.\hskip
  1em plus 0.5em minus 0.4em\relax Berlin, Heidelberg: Springer Berlin
  Heidelberg, 1994, pp. 410--423.

\bibitem{4667215}
H.~Yan, T.~Ren, X.~Peng, X.~Lin, W.~Jiang, T.~Liu, and H.~Guo, ``Information
  reconciliation protocol in quantum key distribution system,'' in \emph{2008
  Fourth International Conference on Natural Computation}, vol.~3, 2008, pp.
  637--641.

\bibitem{zdmw16}
\BIBentryALTinterwordspacing
J.~Zhang, T.~Q. Duong, A.~Marshall, and R.~Woods, ``Key generation from
  wireless channels: A review,'' \emph{IEEE Access}, vol.~4, pp. 614--626,
  2016. [Online]. Available:
  \url{https://dx.doi.org/10.1109/ACCESS.2016.2521718}
\BIBentrySTDinterwordspacing

\bibitem{hamamreh19}
J.~M. Hamamreh, H.~M. Furqan, and H.~Arslan, ``Classifications and applications
  of physical layer security techniques for confidentiality: A comprehensive
  survey,'' \emph{IEEE Communications Surveys \& Tutorials}, vol.~21, no.~2,
  pp. 1773--1828, 2019.

\bibitem{xi2014keep}
W.~Xi, X.-Y. Li, C.~Qian, J.~Han, S.~Tang, J.~Zhao, and K.~Zhao, ``Keep: Fast
  secret key extraction protocol for d2d communication,'' in \emph{2014 IEEE
  22nd International Symposium of Quality of Service (IWQoS)}.\hskip 1em plus
  0.5em minus 0.4em\relax IEEE, 2014, pp. 350--359.

\bibitem{7448884}
J.~Zhang, A.~Marshall, R.~Woods, and T.~Q. Duong, ``Efficient key generation by
  exploiting randomness from channel responses of individual ofdm
  subcarriers,'' \emph{IEEE Transactions on Communications}, vol.~64, no.~6,
  pp. 2578--2588, 2016.

\bibitem{Hern2006:Lightweight}
S.~M. Hernandez and E.~Bulut, ``{Lightweight and Standalone {IoT} Based {WiFi}
  Sensing for Active Repositioning and Mobility},'' in \emph{21st International
  Symposium on {"}A World of Wireless, Mobile and Multimedia Networks{"}
  (WoWMoM 2020)}, Cork, Ireland, Jun. 2020.

\bibitem{9880811}
X.~Yuan, Y.~Jiang, A.~Hu, and C.~Guo, ``Wireless channel key generation for
  multi-user access scenarios,'' in \emph{2022 IEEE/CIC International
  Conference on Communications in China (ICCC)}, 2022, pp. 179--184.

\bibitem{10012875}
G.~Cao, Y.~Zhang, Z.~Ji, M.~Zhang, and Z.~He, ``Esp32-driven physical layer key
  generation: A low-cost, integrated, and portable implementation,'' in
  \emph{2022 IEEE 96th Vehicular Technology Conference (VTC2022-Fall)}, 2022,
  pp. 1--5.

\bibitem{9625573}
C.~Li, Y.~Jiang, and A.~Hu, ``{CSI} measurement and reciprocity evaluation
  method based on embedded platform,'' in \emph{2021 IEEE 94th Vehicular
  Technology Conference (VTC2021-Fall)}, 2021, pp. 1--6.

\bibitem{10.1145/1067170.1067177}
\BIBentryALTinterwordspacing
C.~Castelluccia and P.~Mutaf, ``Shake them up! a movement-based pairing
  protocol for cpu-constrained devices,'' in \emph{Proceedings of the 3rd
  International Conference on Mobile Systems, Applications, and Services}, ser.
  MobiSys '05.\hskip 1em plus 0.5em minus 0.4em\relax New York, NY, USA:
  Association for Computing Machinery, 2005, p. 51–64. [Online]. Available:
  \url{https://doi.org/10.1145/1067170.1067177}
\BIBentrySTDinterwordspacing

\bibitem{10.5555/1771592.1771610}
D.~Bichler, G.~Stromberg, M.~Huemer, and M.~L\"{o}w, ``Key generation based on
  acceleration data of shaking processes,'' in \emph{Proceedings of the 9th
  International Conference on Ubiquitous Computing}, ser. UbiComp '07.\hskip
  1em plus 0.5em minus 0.4em\relax Berlin, Heidelberg: Springer-Verlag, 2007,
  p. 304–317.

\bibitem{4796201}
R.~Mayrhofer and H.~Gellersen, ``Shake well before use: Intuitive and secure
  pairing of mobile devices,'' \emph{IEEE Transactions on Mobile Computing},
  vol.~8, no.~6, pp. 792--806, 2009.

\bibitem{6226870}
Y.~Liu, S.~C. Draper, and A.~M. Sayeed, ``Exploiting channel diversity in
  secret key generation from multipath fading randomness,'' \emph{IEEE
  Transactions on Information Forensics and Security}, vol.~7, no.~5, pp.
  1484--1497, 2012.

\bibitem{9000831}
N.~Aldaghri and H.~Mahdavifar, ``Physical layer secret key generation in static
  environments,'' \emph{IEEE Transactions on Information Forensics and
  Security}, vol.~15, pp. 2692--2705, 2020.

\bibitem{14:MonoPhy}
H.~Abdel-Nasser, R.~Samir, I.~Sabek, and M.~Youssef, ``{MonoPHY}:
  Mono-stream-based device-free {WLAN} localization via physical layer
  information,'' in \emph{WCNC}.\hskip 1em plus 0.5em minus 0.4em\relax IEEE,
  2013, pp. 4546--4551.

\bibitem{2:PinLoc}
S.~Sen, B.~Radunovic, R.~R. Choudhury, and T.~Minka, ``You are facing the {Mona
  Lisa}: Spot localization using {PHY} layer information,'' in
  \emph{Proceedings of the 10th International Conference on Mobile Systems,
  Applications, and Services (MobiSys '12)}.\hskip 1em plus 0.5em minus
  0.4em\relax New York, NY, USA: ACM, 2012, pp. 183--196.

\bibitem{4:FIFS}
J.~Xiao, K.~Wu, Y.~Yi, and L.~M. Ni, ``{FIFS}: Fine-grained indoor
  fingerprinting system,'' in \emph{ICCCN}, 2012, pp. 1--7.

\bibitem{6:FILA}
K.~Wu, J.~Xiao, Y.~Yi, M.~Gao, and L.~M. Ni, ``{FILA}: Fine-grained indoor
  localization,'' in \emph{INFOCOM}, A.~G. Greenberg and K.~Sohraby, Eds.\hskip
  1em plus 0.5em minus 0.4em\relax IEEE, 2012, pp. 2210--2218.

\bibitem{6:CSI}
K.~Wu, J.~Xiao, Y.~Yi, D.~Chen, X.~Luo, and L.~M. Ni, ``{CSI}-based indoor
  localization,'' \emph{{IEEE} Trans. Parallel Distrib. Syst.}, vol.~24, no.~7,
  pp. 1300--1309, 2013.

\bibitem{taw15}
A.~Tzur, O.~Amrani, and A.Wool, ``Direction finding of rogue {Wi-Fi} access
  points using an off-the-shelf {MIMO-OFDM} receiver,'' \emph{Physical
  Communication}, vol.~17, pp. 149--164, Dec. 2015.

\end{thebibliography}



\end{document}